\documentclass[format=sigconf,nonacm=true]{acmart}
\usepackage{tikz}
\usepackage{amsmath}
\usepackage[skip=0.5\baselineskip]{caption}
\usepackage{booktabs}
\usepackage{adjustbox}
\usepackage{multirow}
\usepackage{todonotes}
\usepackage{paralist}
\usepackage{xcolor}
\usepackage{multicol}
\usepackage{float}

\AtBeginDocument{%
  }

\setcopyright{acmcopyright}
\copyrightyear{2022}
\acmYear{2022}
\acmDOI{XXXXXXX.XXXXXXX}

\acmConference[AISec '22]{15th ACM Workshop on Artificial Intelligence and Security}{November 11, 2022}{Los Angeles, USA}
\acmPrice{15.00}
\acmISBN{978-1-4503-XXXX-X/18/06}

\begin{document}

\title{Label-Only Membership Inference Attack against \\Node-Level Graph Neural Networks}




\author{Mauro Conti}
\affiliation{%
 \institution{University of Padua \& Delft University of Technology}
 \country{Italy}}
\email{conti@math.unipd.it}

\author{Jiaxin Li}
\authornote{corresponding author.}
\affiliation{%
  \institution{University of Padua}
  \country{Italy}}
\email{jiaxin.li@studenti.unipd.it}

\author{Stjepan Picek}
\affiliation{%
  \institution{Radboud University \& Delft University of Technology}
  \country{Netherlands}}
\email{stjepan.picek@ru.nl}

\author{Jing Xu}
\affiliation{%
  \institution{Delft University of Technology}
  \country{Netherlands}}
\email{j.xu-8@tudelft.nl}

\renewcommand{\shortauthors}{Conti et al.}

\begin{abstract}
Graph Neural Networks (GNNs), inspired by Convolutional Neural Networks (CNNs), aggregate the message of nodes' neighbors and structure information to acquire expressive representations of nodes for node classification, graph classification, and link prediction. Previous studies have indicated that GNNs are vulnerable to Membership Inference Attacks (MIAs), which infer whether a node is in the training data of GNNs and leak the node's private information, like the patient's disease history. The implementation of previous MIAs takes advantage of the models' probability output, which is infeasible if GNNs only provide the prediction label (label-only) for the input.\\

In this paper, we propose a label-only MIA against GNNs for node classification with the help of GNNs' flexible prediction mechanism, e.g., obtaining the prediction label of one node even when neighbors' information is unavailable. Our attacking method achieves around 60\% accuracy, precision, and Area Under the Curve (AUC) for most datasets and GNN models, some of which are competitive or even better than state-of-the-art probability-based MIAs implemented under our environment and settings. Additionally, we analyze the influence of the sampling method, model selection approach, and overfitting level on the attack performance of our label-only MIA. Both of those factors have an impact on the attack performance. Then, we consider scenarios where assumptions about the adversary's additional dataset (shadow dataset) and extra information about the target model are relaxed. Even in those scenarios, our label-only MIA achieves a better attack performance in most cases. Finally, we explore the effectiveness of possible defenses, including Dropout, Regularization, Normalization, and Jumping knowledge. None of those four defenses prevent our attack completely. 
\end{abstract}



\maketitle

\section{Introduction}
Graph Neural Networks (GNNs) are gaining more and more attention for their broad application in analyzing social networks, recommender systems, and biological networks. Node classification, which predicts the label for nodes in the graph, is one of the popular tasks~\cite{Fan_Graph_2019,Wang_Graph_2019,Hong_reusability_2021}. During the training of GNNs, the model recursively updates nodes' representation from neighbors' message and structure information, similar to the convolution operation of Convolutional Neural Networks (CNNs) on an image. Due to this message passing way, GNNs achieve a competitive performance on multiple tasks~\cite{Zhang_Adaptive_2019,Shang_End_2019}. Still, GNNs are vulnerable to Membership Inference Attacks (MIAs), which infer the existence of nodes in the training data of GNNs. In other words, the adversary can determine which nodes belong to the training data of GNNs with the implementation of MIAs. 

Studies about MIAs against GNNs are dedicated to finding, defending, and understanding this threat. More importantly, in some scenarios, being in the training data, which is nodes' private information, should not be leaked. For instance, when the social network of patients infected with COVID-19 is in the training data of the GNN for investigating infection factors~\cite{olatunji_membership_2021}. Then, if the adversary attacks this GNN with MIAs, the adversary can obtain a list of infected patients. There is no doubt that the history of the disease is the patient's private information. Therefore, studying MIAs against GNNs is essential and significant.

There are two types of MIAs: one is classifier-based MIA, and the other is non-classifier-based MIA. The classifier-based MIA leverages a binary classifier (called attack model), trained with attack features, to infer whether or not one data point is a member of training data. In this paper, the attack features are properties of one data point (or node) for feeding into the attack model. The non-classifier-based MIA directly or indirectly calculates a metric value for one data point and compares this value with a threshold for determining membership. The original intuition of MIA is that the overfitting model will assign a higher probability value to the training data than the testing data. Hence, previous MIAs against GNNs utilize the prediction probability vector to act as attack features or compute metric values. However, they are ineffective when the model only outputs the input's label, which is a label-only condition. Although some label-only MIAs against CNNs and semantic segmentation models exist, label-only MIA against GNNs is still unexplored as far as we know. To rectify it, we propose the label-only MIA against GNNs.

The critical point of our label-only MIA is the attack features extraction for training the attack model. Under the label-only condition, the adversary cannot obtain the prediction probability vector from the target model for the attack feature extraction. Usually, the adversary has access to a shadow dataset from the same distribution as the target dataset, which is used for training and testing the target model. Therefore, the attacker can train a shadow model with the shadow dataset to mimic the behavior of the target model. To improve the effect of imitation, the adversary could relabel the shadow dataset with the target model and train a relabelled shadow model. Consequently, previous label-only works (not on GNNs) acquire the prediction probability vector of the shadow model or the relabelled shadow model as the attack features. 

Unlike previous methods, we observe a flexible prediction mechanism of GNNs, which means that we can obtain the prediction label of a specific node with or without its neighbors' features and connection information fed. Besides, inspired by previous data augmentation methods, we utilize feature masking and step-by-step edge dropping to measure the stability of the prediction label. Specifically, we obtain the attack features of the attack model from three aspects. The first one is the fixed properties of the node, including the number of neighbors and the ground truth of the node. The second one is the prediction correctness of the target model while only feeding the node's features into the GNN and masking the node's properties with different rates and values. The last one is putting the node's features, features of 1-hop neighbors, and connection information into the GNN. Then, the edges between the node and 1-hop neighbors are dropped step by step, and the node's features are masked with different rates and values to get the prediction correctness of the node and its 1-hop neighbors. We obtain the attack features from those three aspects for training our label-only attack model.

Furthermore, we relax the assumptions that the shadow dataset is from the same distribution as the target dataset and that the GNN architecture and type of the shadow model are the same as the target model. Besides, we explore the influence of the sampling ways, model selection strategies, and the overfitting level on the performance of the attack model. Finally, we explore the robustness of our label-only MIA against several possible defenses.

Our main contributions are:
\begin{compactenum}
\item We propose a label-only MIA, which achieves competitive or better performance than state-of-the-art probability-based MIAs.

\item We relax the assumptions about the shadow dataset and target model (the GNN architecture and type), which achieves a better performance in most cases.

\item We explore the influence of the sampling ways, model selection strategies, and the overfitting level on the attack performance. Both of them have an impact on the attack performance. 

\item We analyze the possible defenses against our label-only MIA, where we find that none of the defenses can prevent our attack completely.
\end{compactenum}
To foster reproducibility, we will release the code after the acceptance of the paper.

\section{Background}
\label{sec:background}
We introduce some background knowledge about GNN and MIA for further explanation. In Section~\ref{GNN}, we first review the general GNN architectures. Then, we present the implementation of MIAs in Section~\ref{MIA}. 

\subsection{Graph Neural Networks}
\label{GNN}

GNNs utilize the structure information and nodes' features to update nodes' representations for prediction. In this section, we interpret the basic notions and general architectures of GNNs.   

\subsubsection{Notations}

A graph \textit{$G=(V,E)$} consists of |\textit{$V$}| vertexes and |\textit{$E$}| edges. \textit{$V$} denotes the set of vertexes \{\textit{$V_1,V_2,\ldots,V_{|V|}$}\} with features \textit{$X$}=\{\textit{$X_1,X_2,\ldots,X_{|V|}$}\}. \textit{$E$} represents the set of edges \{\textit{$E_1,E_2,\ldots,E_{|E|}$}\}, each of which connects two nodes in the graph. For node classification tasks, each node \textit{$V_i$} has a label \textit{$Y_i$}. The purpose of node classification tasks is to predict the label of the node with GNNs. GNNs leverage the node's feature and the message passed from the node's neighborhoods to decide its label on the node classification task. We define the \textit{l}-hop neighbors of node \textit{$V_i$} as \textit{$N^l(V_i)$}, which contains a set of nodes whose distance with node \textit{$V_{i}$} is equal to or less than \textit{l} except node \textit{$V_{i}$} itself. For convenience, \textit{$N(V_i)$} denotes 1-hop neighbors of node \textit{$V_{i}$}. Correspondingly, node \textit{$V_{i}$}, its \textit{l}-hop neighbors, connected edges, and relative features make the \textit{l}-hop subgraph, denoted as \textit{$g^l(V_{i})$}, of node \textit{$V_{i}$}.

\subsubsection{General GNN Architecture}

The successful CNN applications empirically prove that the local convolution operation can capture an efficient feature representation of an image for downstream classification tasks. Taking advantage of a similar idea, GNNs aggregate and transfer the message of neighbors into the node's representation, which also compresses the graph structure information. We select four commonly used GNNs, including Graph Convolutional Network (GCN)~\cite{kipf_semi-supervised_2016}, Graph Attention Network (GAT)~\cite{velickovic_graph_2017}, SAmple and aggreGatE (GraphSAGE)~\cite{hamilton_inductive_2017}, and Graph Isomorphism Network (GIN)~\cite{xu_how_2018}. The \textit{AGGREGATION} and \textit{TRANSFORMATION} methods make them distinctive. For each node \textit{$V_{i}$}, \textit{$X_{i}^{(l)}$} is the representation at layer \textit{l} within the iterative convolution, \textit{$H_{i}^{(l)}$} is its hidden state before the \textit{TRANSFORMATION} operation. \textit{$X_{i}^{(0)}$} is the original feature of node \textit{$V_{i}$}. Thus, the general procedures of convolution are defined in the following equations.

\begin{equation}
\label{Aggregate_formula}
\textit{$H_{i}^{(l)}=AGGREGATION(\{X^{(l-1)}_{i}, X^{(l-1)}_{j}, j\in \{N(V_i)\}\})$}.
\end{equation}

\begin{equation}
\label{Transfer_formula}
\textit{$X_{i}^{(l)}=TRANSFORMATION(H_{i}^{(l)})$}.
\end{equation}

Here, \textit{$N(V_i)$} is the 1-hop neighbors of node \textit{$V_{i}$}. The \textit{AGGREGATION} operation clusters the representations of the current node with its neighbors to obtain its presentation. The \textit{TRANSFORMATION} operation is usually a nonlinear conversion. The distinctions of four GNNs are based on those two operations.

\subsection{Membership Inference Attack}
\label{MIA}

The MIA aims to distinguish the training (members) and testing data (non-members) of the target model. Under the context of the node classification, the MIA determines which nodes are in the training data. Generally, there are two types of attack strategies. One is to train a classifier, with which we can predict the possibility of being a member or non-member for a specific data point~\cite{Salem_Membership_2019,shokri_membership_2017}. The other is to directly or indirectly compute the metric about the data point~\cite{li_label-leaks_2020,choquette-choo_label-only_2021,Long_pragmatic_2020}. Then, the attacker determines the data point as a member by comparing the metric value with a threshold. The label-only MIA in our paper belongs to the first type. Here, we provide more details about the classifier-based MIA.

\textbf{Classifier-based MIA.} The key of classifier-based MIA is acquiring the dataset (also called attack dataset) for training the classifier to distinguish the training and testing data. Usually, the adversary has access to a shadow dataset that has the same distribution as the target dataset. With this shadow dataset, the adversary trains a shadow model to mimic the behavior of the target model. Even though the adversary can only query the target model, the adversary has full knowledge of the shadow model. Then, the adversary utilizes the prediction of the shadow model and data split of the shadow model's training and testing data to construct an attack dataset. Finally, the attacker will train an attack model based on the attack dataset for attacking the target model, i.e., distinguishing the training and testing data of the target model. 

\section{Our Label-only MIA}
\label{sec:attack}
GNNs for node classification tasks have two categories, inductive and transductive. Under the inductive setting, the GNN cannot access the testing nodes during training. In the transductive environment, the training set consists of training nodes, unlabelled testing nodes, and nodes' connections. In this paper, we do not consider the transductive setting because the testing data partially acts as the training materials and is not separated from the training data, which mismatches with the knowledge of MIA. He et al.~\cite{he_node-level_2021} also only focused on the inductive setting. While predicting the label of a specific node in the graph, the GNN takes its features, possible neighbors' features, and connection structure information for calculation. However, the GNN could also predict the node's label if only providing the property of a node without neighbors' knowledge. We call this way of prediction the \textit{flexible prediction mechanism}. The flexible way of prediction brings advantages to our label-only MIA, which only gets the label from the model. We formulate our label-only MIA in Section~\ref{problem formulation}. Then, we expose the threat model and attack methodology in Section~\ref{threat model} and Section~\ref{attack methodology}. 

\subsection{Problem Formulation}
\label{problem formulation}

The goal of the MIA is to determine whether the target node \textit{$V_t$} belongs to the training data (member) or testing data (non-member) of the target model \textit{$F_t$}. To implement the MIA, the adversary has some external knowledge \textit{E}. 
We formulate our label-only MIA \textit{A} as the following function:
\begin{equation}
\label{formulation}
\textit{$A:V_t, F_t ,E \rightarrow \{1, 0\}.$}
\end{equation}
Here, "1" means \textit{$V_t$} is in the training data, and "0" otherwise (thus, it is a binary classification task). In the experiments, we train a binary classifier to solve this task.

\subsection{Threat Model}
\label{threat model}

The target model \textit{$F_t$} is open to the adversary, which means the attacker can query it for the label of the target node \textit{$V_t$}. Considering flexible prediction mechanism and practical situation, we limit the query information to the target node's feature, features of its 1-hop neighbors \textit{$N(V_t)$}, and connection structure \textit{$C(V_t)$}. There are two reasons for selecting 1-hop neighbors. The first one is that it requires less information than 2-hop neighbors~\cite{he_node-level_2021} and all the training or testing data~\cite{olatunji_membership_2021} while predicting the membership of one node. The second reason is that the information of the target node is finite under the label-only condition. We cannot get enough helpful distinguishable signals without neighbors' information. Therefore, we choose 1-hop neighbors. Besides, we expose the target node's true label \textit{$Y_{V_t}$} to the adversary.

The adversary has some external knowledge \textit{$E$}. A shadow dataset \textit{$D_s$}, extracted from the same distribution as the target dataset \textit{$D_t$} used for training and testing the target model \textit{$F_t$}, is in the external knowledge. We relax this assumption and sample the shadow dataset from the other distribution in experiments. Furthermore, the adversary knows the target model's architecture, hyperparameters, and training algorithm. With that knowledge, the attacker can train a shadow model \textit{$F_s$} to mimic the behavior of the target model. Similarly, we relax the assumption that the adversary knows the GNN architecture and type of the target model within exploration. The relaxation of those two assumptions tests the limits of our label-only MIA under stricter conditions and is closer to the attack in the real world. He et al.~\cite{he_node-level_2021} and Olatunji et al.~\cite{olatunji_membership_2021} also relaxed those assumptions.

\subsection{Attack Methodology}
\label{attack methodology}
The adversary obtains some external knowledge for the MIA implementation. Then the question is how to apply our label-only MIA on GNNs with that knowledge and still satisfy the threat model. To answer this question, we formulate the final attack model as a binary classifier to predict whether the node is in the training data of the GNN or not. Therefore, the acquisition of the attack features for training the attack model is the crucial point.

\subsubsection{General Steps}

\begin{figure}[ht]
\centering
\includegraphics[scale=0.27]{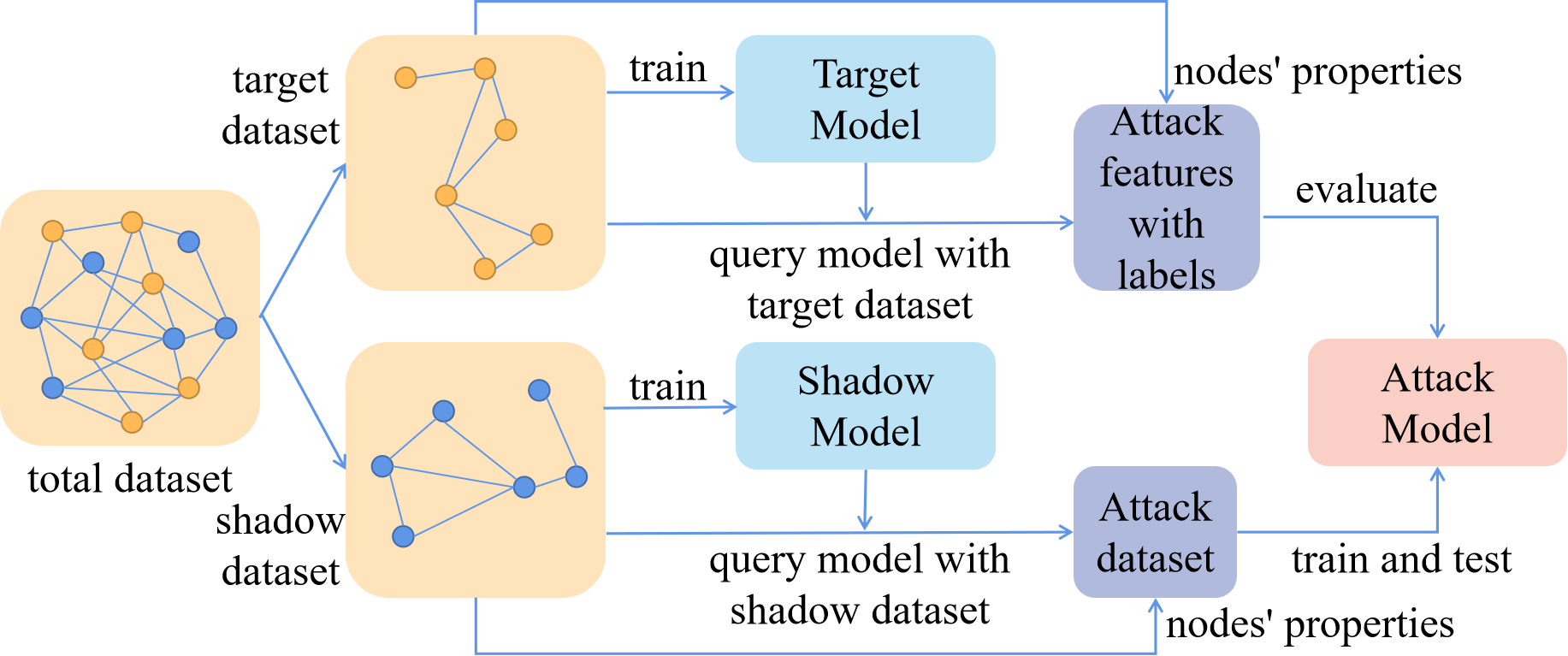}
\caption{The general process of our label-only MIA.}
\label{overall process}
\end{figure}

The overall process is illustrated in Figure~\ref{overall process}. First, the total dataset is split into target and shadow datasets. With a shadow dataset and knowledge about the training of the target model, the adversary can train a shadow model to mimic the behavior of the target model. The shadow model and membership situation of nodes in the shadow dataset are transparent to the adversary. Thus, the attacker generates the attack dataset via data points' fixed properties and multiple queries for each data point to the shadow model, which only outputs the prediction label. The features of the data point in the attack dataset are attack features. If the node is in the shadow model's training data, the adversary assigns its attack features with the label "1" and otherwise "0". Then, the adversary trains the attack model with the attack dataset extracted from the shadow dataset and model. 

Similarly, the adversary queries the target model with target nodes from the target dataset to get each target node's attack features with labels, which indicate whether nodes are from the target model's training data ("1") or testing data ("0"). Afterward, the attacker feeds the target nodes' attack features into the attack model to test the attack performance on the target model. Notably, the acquisition of the attack features is interpreted in the next section. 

\subsubsection{Attack Features Acquisition}
\label{feature acquisition}

Under the label-only condition, the adversary only gets the prediction label of the target node from the target model. Therefore, obtaining the attack features of target nodes in the target dataset is crucial. Previous label-only MIAs~\cite{li_label-leaks_2020} (not on GNNs) extract attack features of the target node from the prediction probability vector of the shadow or relabelled shadow models, which are under the adversary's control. The relabelled shadow model utilizes the target model to relabel the shadow dataset, which is different from the shadow model. Previous methods feed the target node into shadow or relabelled shadow models to get the prediction probability vector as the attack features of this node under the label-only condition because the target model does not expose the prediction probability vector.



Unlike previous methods, we construct target nodes' attack features based on fixed properties of the node, data augmentation, flexible prediction mechanism, and practical situation. The application of data augmentation obtains inspiration from previous works~\cite{choquette-choo_label-only_2021,Ding_augmentation_2022}. We consider the attack features from three aspects: fixed properties, 0-hop query, and 1-hop query. The specified properties, like the ground truth of the target node in the target model, act as a part of features in previous attack models~\cite{shokri_membership_2017,Hui_Blind_2021}. Besides, the adversary has only access to 1-hop neighbors, which means the attacker could query the target model with (2-hop query) or without (0-hop query) information of 1-hop neighbors. Therefore, we extract features from the three mentioned aspects.

\textbf{Acquisition with fixed properties.} Olatunji et al.~\cite{olatunji_membership_2021} mentioned that the connection between nodes increases the vulnerability of GNN models to privacy attacks. Inspired by this, we count the number of neighbors (n\_num) and signal (w\_i\_node), which indicates whether the node is independent of other nodes, as part of the attack features of the target node \textit{$V_t$}. Notably, the neighbors could be in the training or testing data of the dataset while counting the number of neighbors in the target or shadow dataset. Besides, we include the target node's ground truth (o\_label) into the attack features, which is the same as the previous MIA~\cite{shokri_membership_2017}.

\textbf{Acquisition with the 0-hop query.} The prediction of GNNs is flexible, which means we can get the prediction label of the node with or without the features and the connection information of its neighbors fed into GNNs. For each target node \textit{$V_t$}, we randomly change different rates of its feature values \textit{$X_{V_t}$} to the largest and smallest value in the feature space. Then we only input this changed feature of the target node to the target model and record whether the prediction is the same as the ground truth of the target node. Here, \textit{$i\_none\_max\_rate$} and \textit{$i\_none\_min\_rate$} are the record results for each \textit{rate} in \textit{$rate\_set$}. The "1" means the prediction of the changed feature is the same as the ground truth of the target node and the "0" otherwise.

\textbf{Acquisition with the 1-hop query.} Due to the flexible prediction mechanism of GNNs, we feed the features \textit{$X_{V_t}$} of the target node \textit{$V_t$}, the features \textit{$X_{N(V_t)}$} of its 1-hop neighbors \textit{$N(V_t)$}, and connection information \textit{$C(V_t)$} between the target node and its 1-hop neighbors into the target model. At the same time, we apply two data augmentation strategies simultaneously. One is randomly changing different rates of feature values to the largest and smallest value in the feature space. The other is dropping the connected edges between the target node and its 1-hop neighbors one by one. Then, we record the prediction accuracy of the 1-hop neighbors and the identity between the prediction label and the target node's ground truth. 

Following the previous paragraph, we explain attack features' names while acquiring with the 1-hop query. Figure~\ref{SHAP_values} consists of features that have a high absolute SHAP value. The \textit{$n\_acc\_all\_max\_rate$} means the accuracy of the neighbors while keeping all the edges and altering a percentage (\textit{rate}) of the target node's feature values to the maximum value in the feature space. The \textit{$n\_acc\_none\_max\_rate$} is similar except all the edges are removed. The \textit{$n\_acc\_avg\_max\_rate$} is the average accuracy of the neighbors during removing edges. The \textit{$i\_all\_max\_rate$} indicates whether the prediction of the target node's changed feature is the same as its ground truth while keeping all the edges. The \textit{$i\_step\_max\_rate$} presents the rate of cases where the prediction of the target node's changed feature is the same as its ground truth while reducing edges step by step. The names of attack features are slightly different while changing the feature value to the smallest value, i.e., replacing "max" with "min" in the feature name. Besides, we record the real change percentage \textit{$change\_p\_rate$} of the feature values while randomly select a percentage (\textit{rate}) of feature values for altering.

Our label-only MIA combines fixed properties, the result of the 0-hop query, and the result of the 1-hop query as the attack features. Previous studies~\cite{he_node-level_2021,olatunji_membership_2021,duddu_2020_quantify} obtained the prediction of the target node with 2-hop neighbors or all nodes in the training or testing data as the input of the target model. They contains more information than our label-only MIA while predicting the node's label. Under the label-only condition, we cannot get the prediction probability vector from the target model. However, while dropping edges or altering its feature values, the target node and its neighbors' prediction correctness situation could also provide the signal for distinguishing the training and testing data.

\section{Experiments}
\label{sec:setup}
We first describe datasets in our experiments, followed by describing models' architectures and training settings. In Section~\ref{evaluation metrics}, we introduce evaluation metrics. Finally, we illustrate experimental steps.

\subsection{Datasets}

We conduct experiments on four datasets: Cora\_ML, CiteSeer, DBLP, and PubMed~\cite{Aleksandar_Gaussian_2018}. The main reason for selecting those datasets is that the number of nodes and classes of those datasets varies, which is beneficial for various experiments. Table~\ref{dataset_infor} describes the statistics of those four datasets.

\begin{table}[ht]
\centering
\caption{Statistical information of datasets.}
\label{dataset_infor}
\begin{adjustbox}{max width=0.47\textwidth}
\begin{tabular}{c|cccc}
\toprule
Dataset  & classes & edges  & nodes & length of node feature \\
\midrule
Cora\_ML & 7       & 16,316  & 2,995  & 2,879                   \\
CiteSeer & 6       & 10,674  & 4,230  & 602                    \\
DBLP     & 4       & 105,734 & 17,716 & 1,639                   \\
PubMed   & 3       & 88,648  & 19,717 & 500   \\                
\bottomrule
\end{tabular}
\end{adjustbox}
\end{table}

\subsection{Model Architectures and Training Settings}

The GNN types we consider are GCN, GAT, GraphSAGE, and GIN. We vary the model architectures and training settings to explore the influence of the overfitting level on the attack performance. Here, we discuss model architectures and training settings for target models with high and low overfitting levels.

\textbf{Low overfitting level.} In this setting, we set four types of GNNs with three layers (input, hidden, and output layers), with 16 neurons in the hidden layer. The optimization algorithm of GNNs is Adam, with a learning rate of 6e-3 and a weight decay of 0.5. The number of training epochs is 400. Besides, we apply the BatchNorm, Dropout (0.5), and Jumping knowledge (concatenation)~\cite{Xu_jump_2018} to reduce the overfitting level. Jumping knowledge (concatenation) concatenates the output of each layer as the input of the last layer.

\textbf{High overfitting level.} Different from models with low overfitting levels, we set the number of layers to 5 (3 hidden layers) and the number of neurons in the hidden layer to 64. The optimization algorithm of GNNs is Adam, with a learning rate of 0.001 and without weight decay. The number of training epochs is 200. In addition, we do not apply any strategies to reduce the overfitting level because we attempt to get the model with a high overfitting level for comparison.

We decide on the training hyperparameters of models with a high or low overfitting level by increasing or decreasing the overfitting level in a reasonable range. Besides, we refer to the settings in previous works~\cite{he_node-level_2021,olatunji_membership_2021}.

The attack model is a multilayer perceptron (MLP) with 2 hidden layers, each of which has 64 neurons. The optimization algorithm is Adam, with a learning rate of 0.001. The number of training epochs is 300, and the batch size is 32. We use the Binary Cross Entropy to guide the training of the attack model. Our attack model's architecture is similar to attack models in previous works~\cite{he_node-level_2021,olatunji_membership_2021}. After several attempts, we slightly changed the structure and hyperparameters to fit attack features in our label-only MIA. Finally, we obtained the attack model's structure and hyperparameters mentioned before.

\subsection{Evaluation Metrics}
\label{evaluation metrics}

We train the attack model with the objective function related to the Binary Cross Entropy. The output of the attack model is the probability of the input to be a member of training data. We evaluate the attack performance with six metrics: precision, recall, F1 while the threshold is 0.5, accuracy, AUC, and True Positive Rate (TPR) while the False Positive Rate is 0.1, which is inspired by the work of Carlini et al.~\cite{Carlini_Membership_2022}.

\subsection{Experimental Steps}
\label{experiment steps}

\begin{table*}[ht]
\centering
\caption{The attack performance of our label-only MIA after ten repetitions (low overfitting level).}
\label{label-only MIA performance}
\begin{adjustbox}{max width=0.75\textwidth}
\begin{tabular}{cccccc}
\hline
\multirow{2}{*}{Dataset}  & \multirow{2}{*}{GNN} & \multirow{2}{*}{Used Nodes} & \multirow{2}{*}{\begin{tabular}[c]{@{}c@{}}Test Acc\\ (target model)\end{tabular}} & \multirow{2}{*}{\begin{tabular}[c]{@{}c@{}}Train Acc\\ (target model)\end{tabular}} & \multirow{2}{*}{\begin{tabular}[c]{@{}c@{}}Our Attack\\ (avg acc, pre, rec, auc, f1, low\_fpr\_0.01\_tpr)\end{tabular}} \\
                          &                      &                             &                                                                                     &                                                                                      &                                                                                                                         \\ \hline
\multirow{4}{*}{Cora\_ML} & GAT                  & \multirow{4}{*}{1,344}       & 0.736                                                                               & 0.883                                                                                & {[}\textcolor{red}{0.604}, 0.605, 0.618, 0.658, 0.607, 0.059{]}                                                                          \\ \cline{6-6} 
                          & GCN                  &                             & 0.763                                                                               & 0.951                                                                                & {[}\textcolor{red}{0.613}, 0.621, 0.606, 0.666, 0.607, 0.065{]}                                                                          \\ \cline{6-6} 
                          & GIN                  &                             & 0.664                                                                               & 0.875                                                                                & {[}\textcolor{red}{0.59}, 0.624, 0.49, 0.631, 0.536, 0.041{]}                                                                            \\ \cline{6-6} 
                          & GraphSAGE            &                             & 0.73                                                                                & 0.956                                                                                & {[}\textcolor{red}{0.602}, 0.618, 0.559, 0.65, 0.581, 0.061{]}                                                                           \\ \hline
\multirow{4}{*}{CiteSeer} & GAT                  & \multirow{4}{*}{3,336}       & 0.793                                                                               & 0.879                                                                                & {[}\textcolor{red}{0.598}, 0.611, 0.585, 0.647, 0.583, 0.032{]}                                                                          \\ \cline{6-6} 
                          & GCN                  &                             & 0.81                                                                                & 0.942                                                                                & {[}\textcolor{red}{0.599}, 0.622, 0.536, 0.645, 0.569, 0.031{]}                                                                          \\ \cline{6-6} 
                          & GIN                  &                             & 0.777                                                                               & 0.921                                                                                & {[}\textcolor{red}{0.592}, 0.605, 0.563, 0.635, 0.571, 0.032{]}                                                                          \\ \cline{6-6} 
                          & GraphSAGE            &                             & 0.795                                                                               & 0.937                                                                                & {[}\textcolor{red}{0.581}, 0.571, 0.69, 0.617, 0.621, 0.032{]}                                                                           \\ \hline
\multirow{4}{*}{DBLP}     & GAT                  & \multirow{4}{*}{7,920}       & 0.732                                                                               & 0.83                                                                                 & {[}\textcolor{red}{0.593}, 0.604, 0.586, 0.645, 0.584, 0.048{]}                                                                          \\ \cline{6-6} 
                          & GCN                  &                             & 0.746                                                                               & 0.884                                                                                & {[}\textcolor{red}{0.593}, 0.598, 0.618, 0.644, 0.596, 0.043{]}                                                                          \\ \cline{6-6} 
                          & GIN                  &                             & 0.715                                                                               & 0.865                                                                                & {[}\textcolor{red}{0.579}, 0.59, 0.563, 0.621, 0.564, 0.026{]}                                                                           \\ \cline{6-6} 
                          & GraphSAGE            &                             & 0.739                                                                               & 0.891                                                                                & {[}\textcolor{red}{0.579}, 0.578, 0.643, 0.618, 0.595, 0.032{]}                                                                          \\ \hline
\multirow{4}{*}{PubMed}   & GAT                  & \multirow{4}{*}{12,300}      & 0.861                                                                               & 0.884                                                                                & {[}\textcolor{red}{0.583}, 0.573, 0.661, 0.636, 0.61, 0.033{]}                                                                           \\ \cline{6-6} 
                          & GCN                  &                             & 0.867                                                                               & 0.907                                                                                & {[}\textcolor{red}{0.596}, 0.601, 0.58, 0.657, 0.589, 0.059{]}                                                                           \\ \cline{6-6} 
                          & GIN                  &                             & 0.852                                                                               & 0.899                                                                                & {[}\textcolor{red}{0.568}, 0.572, 0.571, 0.611, 0.563, 0.034{]}                                                                          \\ \cline{6-6} 
                          & GraphSAGE            &                             & 0.866                                                                               & 0.923                                                                                & {[}\textcolor{red}{0.557}, 0.548, 0.665, 0.597, 0.598, 0.033{]}                                                                          \\ \hline
\end{tabular}
\end{adjustbox}
\end{table*}

The implementation of our label-only MIA is composed of several steps. First, we divide the total dataset into training and testing data for target and shadow models, i.e., four sub-datasets. Then, we generate the attack dataset from the shadow model and shadow dataset for training and testing the attack model. After extracting features and labels from the target model and target dataset, we evaluate the attack performance of the attack model on those features and labels. Apart from implementing our label-only MIA, we also implement previous probability-based MIAs under the same environment and settings. Those settings consist of the same models trained with the same sub-datasets, the training of the attack models, and the model selection strategy. Besides, we explore the impact of several factors on our label-only MIA. Those factors include the overfitting level, sampling methods, and attack model selection strategies. Furthermore, we attempt to relax two assumptions of training the shadow model. Finally, we analyze the effectiveness of some defenses.

\textbf{Previous Probability-based MIAs.} To compare our label-only MIA, we implement four probability-based MIAs proposed in previous papers~\cite{he_node-level_2021,olatunji_membership_2021}. Those methods can obtain the prediction probability vector from the target model and extract the attack features from the prediction probability vector, which is impossible under the label-only condition. Then, they train the attack model based on the attack features. Specifically, those four probability-based MIAs use the top two probability values of the 0-hop query or the 2-hop query, the combination of the top two values from 0-hop and 2-hop queries, or all probability values while feeding features of the training or testing data into the model for predicting the label of one node. In our paper, we name four probability-based MIAs: 0-hop, 2-hop, the combination of 0-hop and 2-hop, and all probability methods.

\textbf{Overfitting.} As mentioned in the model architectures and settings, we attempt to explore the influence of the overfitting level on the attack performance. Thus, we train models of different architectures with various settings to expose the impact of the overfitting level on our label-only MIA. 

\textbf{Sampling Methods.} We focus on the node-level GNN in this paper. The dataset for training GNNs is composed of a graph with many nodes and edges. As we mentioned, we need to sample four sub-datasets with equal or approximately equal data points in the total dataset. However, the number of nodes in each class is not equal and even has a large gap. Therefore, the sampling cannot guarantee that each class has an equal number of data points in each sub-dataset unless decreasing the number of data points in each sub-dataset. Based on this situation, we take three sampling methods into comparison. The first is a random sampling of four sub-datasets with maximum utilization of data points (called \textit{the random sampling method}). The second one is a strict class-balanced sampling approach (called \textit{the balanced sampling method}), which guarantees non-overlapping, class-balanced, and fewer data points in each sub-dataset. The third one is \textit{the partially balanced sampling method}, which ensures the class balance in the training data of target and shadow models and randomly samples data points for the testing data.

\textbf{Attack Model Selection Strategies.} We train and test the attack model based on the attack dataset. Then, we evaluate the attack performance of the attack model with attack features. The selection of the attack model during the training process impacts the final attack model. Here, we choose the attack model based on five metrics, including training accuracy (\textit{train\_acc}), testing accuracy (\textit{test\_acc}), training loss (\textit{train\_loss}), testing loss (\textit{test\_loss}), and evaluation accuracy (\textit{evaluate\_acc}). Importantly, the adversary cannot compute evaluation accuracy in a practical scenario because the adversary cannot obtain the attack features of the target dataset for evaluation before selecting the final attack model. We implement the selection strategy based on evaluation accuracy. Finally, we analyze the attack performance of those model selection strategies.

\textbf{Assumptions Relaxation.} The assumptions about the shadow dataset and the target model's information (GNN type and architecture) are not always available in real-world settings. Therefore, we relax those two assumptions alone and together to assess the attack performance of our label-only MIA.

\textbf{Defenses.} The methods for reducing the overfitting level are commonly used to prevent MIAs. To evaluate the effectiveness of our label-only MIA, we apply the Dropout, Regularization, Normalization, and Jumping knowledge to reduce the overfitting level.

\section{Results and Discussions}
\label{sec:results}
In this section, we provide the results of our experiments and discuss the findings from the results. Section~\ref{attack performance comparison} compares the attack performance of our label-only MIA with four previous probability-based MIAs. We interpret the attack model with SHAP values in Section~\ref{attack model explanation}. In Section~\ref{influence factors}, we explore the influence of different factors on the attack performance of the attack model. We relax two assumptions about the shadow dataset and the target model's information in Section~\ref{assumptions relaxation}. Finally, we present the effectiveness of our MIA against four defenses.

\subsection{Attack Performance Comparison}
\label{attack performance comparison}


As mentioned, the adversary trains the shadow model, which has the same architecture as the target model, with the non-overlapping shadow dataset from the same distribution as the target dataset. Table~\ref{label-only MIA performance} provides the attack performance of our label-only MIA after ten repetitions. Repeating each experiment ten times reduces the impact of the randomness, which is the same as the repetition times in the previous works~\cite{olatunji_membership_2021}. 
In the table, the number of data points used in each experiment is related to the balanced sampling method. The overfitting level, measured by the gap between the training and testing accuracy of the target model, is relatively low. From the table, we can see that the attack accuracy, precision, and AUC value are around 0.6 in most experiments, indicating the effectiveness of our label-only MIA against GNNs.

Table~\ref{previous MIA performance} in Appendix~\ref{table_explanation} shows the corresponding attack performance of previous probability-based MIAs under the same environment as Table~\ref{label-only MIA performance}. We highlight the average attack accuracy of each table in red color. Comparing the average attack accuracy of each row in two tables, we can see that our label-only MIA has higher average accuracy than previous probability-based MIA in most cases. For instance, our label-only MIA achieves the average attack accuracy of 0.596 while four previous probability-based MIAs obtain values of 0.506, 0.506, 0.504, and 0.507 of GCN on PubMed dataset. Previous probability-based MIAs only achieve larger accuracies in three models, i.e., GCN, GIN, and GraphSAGE, on Cora\_ML dataset. For example, previous MIAs have average attack accuracies of 0.691, 0.633, 0.693, and 0.635, while our label-only MIA obtains the value of 0.613 under Cora\_ML and GCN. It reflects that our label-only MIA has a competitive, even better performance than previous probability-based methods in most experiments under our environment and settings.

\subsection{Attack Model Explanation}
\label{attack model explanation}

\begin{figure*}[ht]
\centering
\includegraphics[scale=0.45]{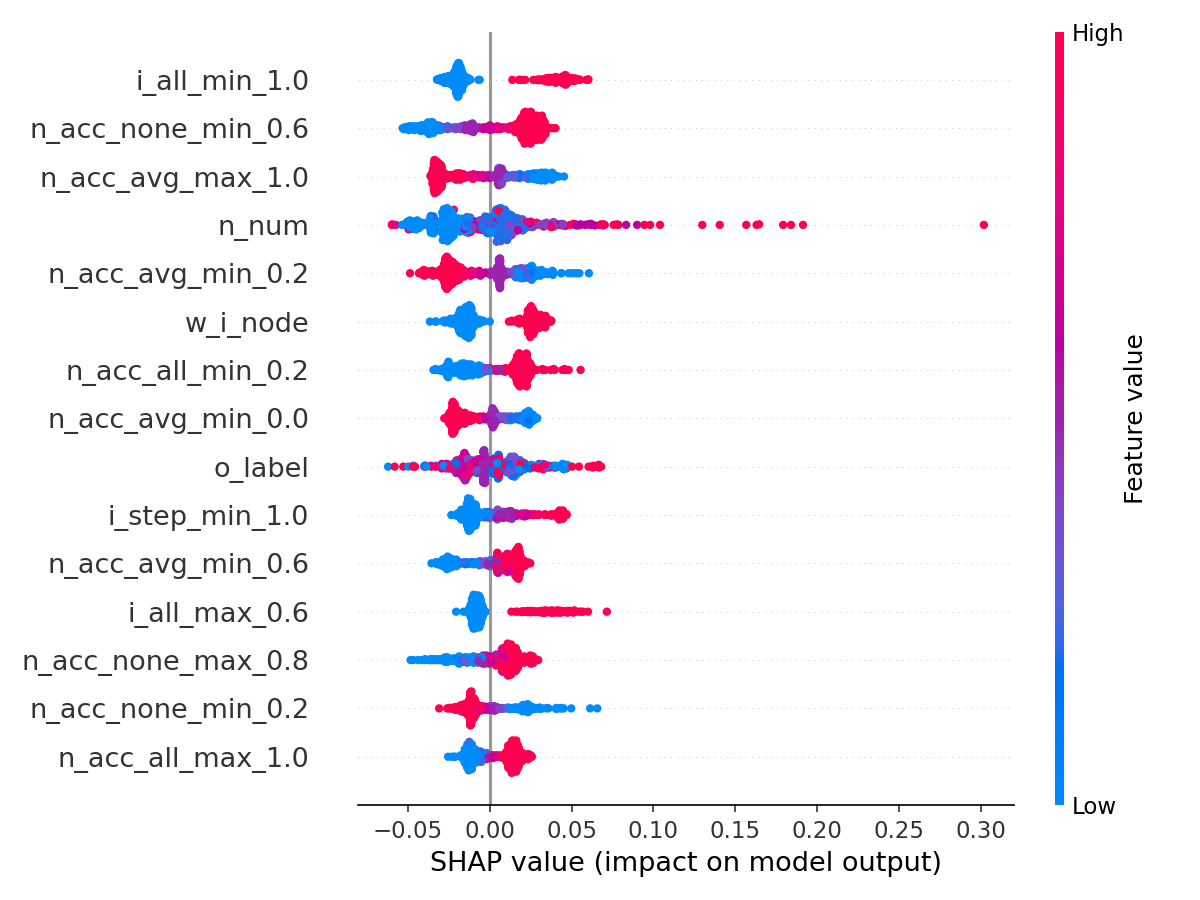}
\caption{The SHAP values of attack features (top 15). The left column indicates the attack features ranked by the absolute SHAP value. The colorful right line represents the ruler of features' values. In the middle, the cluster of points reflects data points in the dataset. The x-axis shows the SHAP value of each feature within one data point.}
\label{SHAP_values}
\end{figure*}

We describe the details of acquiring the features for training the attack model in Section~\ref{feature acquisition}. To better understand the model's behavior and attempt to explain the model, we calculate the SHAP~\cite{lundberg_unified_2017} value of each feature. The SHAP value implies the contribution or importance of a feature to the prediction of the model. Figure~\ref{SHAP_values} gives the SHAP values of each feature in the attack model under the Cora\_ML and GCN setting, which has a higher attack performance for clearly exposing the importance of each feature.

The left column in the figure displays features' names, ranked by the absolute SHAP values of each feature over total data points. In the middle, a large number of colorful points represent total data points with different feature values and SHAP values in each row. The x-axis means the SHAP value. The right line implies the feature values for spots in the middle. We can see that the feature named "i\_all\_min\_1.0" has the top absolute SHAP value. This feature presents whether the prediction of the target model to the target node is the same as the ground truth while keeping all the edges between the target node with 1-hop neighbors and changing 100\% of the target node's features to the minimum value. In addition, we can observe that most features have an apparent positive or negative impact on model output while being assigned high or low values. Besides, the features with the prefix "n\_acc" have higher SHAP values than other features, which exposes that neighboring nodes' accuracy under various settings is beneficial for distinguishing the training and testing data.

\subsection{Influence Factors}
\label{influence factors}

This section explores our label-only MIA's influence factors, including the overfitting level, sampling methods, and attack model selection strategies.

\textbf{Overfitting.} Table~\ref{label-only MIA performace_2} shows the attack performance of our label-only MIA against target models with a high overfitting level. We compare the results in this table with Table~\ref{label-only MIA performance} and highlight the average attack accuracy in the red color. The average attack accuracy could increase as the enlargement of the overfitting level—for example, the DBLP on GAT and GCN models, from 0.593 and 0.593 (low overfitting level) to 0.6 and 0.599 (high overfitting level). On the contrary, the increase in the overfitting level could also reduce our label-only MIA's attack performance. For instance, the average attack accuracies move from high (0.604, 0.613, 0.59, and 0.602) to low values (0.597, 0.596, 0.536, and 0.593) under combinations of Cora\_ML with GAT, GCN, GIN, and GraphSAGE, respectively. Therefore, the higher overfitting level impacts the attack performance, i.e., it could increase or decrease our label-only MIA's attack accuracy. We use the models with a low overfitting level to complete the subsequent exploration.

\begin{table}[ht]
\centering
\caption{The attack performance of our label-only MIA after ten repetitions (high overfitting level).}
\label{label-only MIA performace_2}
\begin{adjustbox}{max width=0.47\textwidth}
\begin{tabular}{ccccc}
\hline
Dataset                   & GNN       & \begin{tabular}[c]{@{}c@{}}Test Acc \\ (target \\ model)\end{tabular} & \begin{tabular}[c]{@{}c@{}}Train Acc\\ (target \\ model)\end{tabular} & \begin{tabular}[c]{@{}c@{}}Our Attack\\ (avg acc, pre, rec, auc, \\ f1, low\_fpr\_0.01\_tpr)\end{tabular} \\ \hline
\multirow{4}{*}{Cora\_ML} & GAT       & 0.658                                                                  & 0.973                                                                  & \begin{tabular}[c]{@{}c@{}}{[}\textcolor{red}{0.597}, 0.635, 0.504, \\ 0.652, 0.545, 0.056{]}\end{tabular}                 \\ \cline{5-5} 
                          & GCN       & 0.684                                                                  & 0.96                                                                   & \begin{tabular}[c]{@{}c@{}}{[}\textcolor{red}{0.596}, 0.605, 0.586, \\ 0.634, 0.586, 0.033{]}\end{tabular}                 \\ \cline{5-5} 
                          & GIN       & 0.298                                                                  & 0.577                                                                  & \begin{tabular}[c]{@{}c@{}}{[}\textcolor{red}{0.536}, 0.539, 0.525, \\ 0.554, 0.525, 0.022{]}\end{tabular}                 \\ \cline{5-5} 
                          & GraphSAGE & 0.619                                                                  & 0.999                                                                  & \begin{tabular}[c]{@{}c@{}}{[}\textcolor{red}{0.593}, 0.602, 0.595, \\ 0.637, 0.59, 0.054{]}\end{tabular}                  \\ \hline
\multirow{4}{*}{CiteSeer} & GAT       & 0.748                                                                  & 0.913                                                                  & \begin{tabular}[c]{@{}c@{}}{[}\textcolor{red}{0.592}, 0.612, 0.546, \\ 0.641, 0.567, 0.042{]}\end{tabular}                 \\ \cline{5-5} 
                          & GCN       & 0.767                                                                  & 0.922                                                                  & \begin{tabular}[c]{@{}c@{}}{[}\textcolor{red}{0.603}, 0.62, 0.572, \\ 0.656, 0.586, 0.043{]}\end{tabular}                  \\ \cline{5-5} 
                          & GIN       & 0.327                                                                  & 0.578                                                                  & \begin{tabular}[c]{@{}c@{}}{[}\textcolor{red}{0.538}, 0.536, 0.569, \\ 0.556, 0.549, 0.029{]}\end{tabular}                 \\ \cline{5-5} 
                          & GraphSAGE & 0.732                                                                  & 0.874                                                                  & \begin{tabular}[c]{@{}c@{}}{[}\textcolor{red}{0.591}, 0.618, 0.532, \\ 0.637, 0.557, 0.027{]}\end{tabular}                 \\ \hline
\multirow{4}{*}{DBLP}     & GAT       & 0.716                                                                  & 0.916                                                                  & \begin{tabular}[c]{@{}c@{}}{[}\textcolor{red}{0.6}, 0.617, 0.585, \\ 0.648, 0.582, 0.044{]}\end{tabular}                   \\ \cline{5-5} 
                          & GCN       & 0.722                                                                  & 0.917                                                                  & \begin{tabular}[c]{@{}c@{}}{[}\textcolor{red}{0.599}, 0.627, 0.549, \\ 0.653, 0.572, 0.062{]}\end{tabular}                 \\ \cline{5-5} 
                          & GIN       & 0.428                                                                  & 0.652                                                                  & \begin{tabular}[c]{@{}c@{}}{[}\textcolor{red}{0.555}, 0.561, 0.509, \\ 0.583, 0.522, 0.025{]}\end{tabular}                 \\ \cline{5-5} 
                          & GraphSAGE & 0.666                                                                  & 0.856                                                                  & \begin{tabular}[c]{@{}c@{}}{[}\textcolor{red}{0.579}, 0.586, 0.593, \\ 0.618, 0.577, 0.029{]}\end{tabular}                 \\ \hline
\multirow{4}{*}{PubMed}   & GAT       & 0.828                                                                  & 0.912                                                                  & \begin{tabular}[c]{@{}c@{}}{[}\textcolor{red}{0.58}, 0.616, 0.477, \\ 0.62, 0.524, 0.033{]}\end{tabular}                   \\ \cline{5-5} 
                          & GCN       & 0.838                                                                  & 0.909                                                                  & \begin{tabular}[c]{@{}c@{}}{[}\textcolor{red}{0.596}, 0.611, 0.561, \\ 0.653, 0.575, 0.058{]}\end{tabular}                 \\ \cline{5-5} 
                          & GIN       & 0.586                                                                  & 0.639                                                                  & \begin{tabular}[c]{@{}c@{}}{[}\textcolor{red}{0.522}, 0.52, 0.607, \\ 0.537, 0.542, 0.018{]}\end{tabular}                  \\ \cline{5-5} 
                          & GraphSAGE & 0.836                                                                  & 0.921                                                                  & \begin{tabular}[c]{@{}c@{}}{[}\textcolor{red}{0.561}, 0.558, 0.672, \\ 0.594, 0.603, 0.031{]}\end{tabular}                 \\ \hline
\end{tabular}
\end{adjustbox}
\end{table}

\textbf{Sampling Methods.} As mentioned in Section~\ref{experiment steps}, we explore the impact of three sampling methods, including the (0) random, (1) balanced, and (2) partially balanced sampling methods. Table~\ref{data points comparison} exposes the number of data points in four sub-datasets, including the training and testing data of shadow and target models under different sampling methods. In the table, target\_train, target\_test, shadow\_train, and shadow\_test represent four sub-datasets. The "total" means the number of data points in this sub-dataset. The "one class" indicates the number of data points in each class of this sub-dataset. The "-" means uncertainty indicating that the number of data points in each class is not fixed due to random selection. From the table, the uncertainty represents the class imbalance in the random and partially balanced sampling methods.

\begin{table}[ht]
\centering
\caption{The number of data points in each sub-dataset under different sampling methods.}
\label{data points comparison}
\begin{adjustbox}{max width=0.48\textwidth}
\begin{tabular}{cccccccccc}
\hline
\multirow{5}{*}{\rotatebox{90}{Dataset}}  & \multirow{5}{*}{\begin{tabular}[c]{@{}c@{}}Sampling\\ Method\end{tabular}} & \multicolumn{2}{c}{target\_train} & \multicolumn{2}{c}{target\_test} & \multicolumn{2}{c}{shadow\_train} & \multicolumn{2}{c}{shadow\_test} \\ \cline{3-10} 
                          &                                                                & \rotatebox{90}{total}         & \rotatebox{90}{one class}         & \rotatebox{90}{total}         & \rotatebox{90}{one class}        & \rotatebox{90}{total}         & \rotatebox{90}{one class}         & \rotatebox{90}{total}         & \rotatebox{90}{one class}        \\ \hline
\multirow{3}{*}{\rotatebox{90}{Cora\_ML}} & 0                                                                          & 749           & -                 & 749           & -                & 748           & -                 & 749           & -                \\
                          & 1                                                                          & 336           & 48                & 336           & 48               & 336           & 48                & 336           & 48               \\
                          & 2                                                                          & 630           & 90                & 630           & -                & 630           & 90                & 630           & -                \\ \hline
\multirow{3}{*}{\rotatebox{90}{CiteSeer}} & 0                                                                          & 1,058           & -                 & 1,057           & -                & 1,058           & -                 & 1,057           & -                \\
                          & 1                                                                          & 834           & 139                & 834           & 139               & 834           & 139               & 834           & 139               \\
                          & 2                                                                          & 600           & 100                & 600           & -                & 600           & 100                & 600           & -                \\ \hline
\multirow{3}{*}{\rotatebox{90}{DBLP}} & 0                                                                          & 4,429           & -                 & 4,429           & -                & 4,429          & -                 & 4,429           & -                \\
                          & 1                                                                          & 1,980           & 495                & 1,980           & 495               & 1,980           & 495               & 1,980           & 495               \\
                          & 2                                                                          & 3,200           & 800                & 3,200           & -                & 3,200           & 800                & 3,200           & -                \\ \hline
\multirow{3}{*}{\rotatebox{90}{PubMed}} & 0                                                                          & 4,929           & -                 & 4,929           & -                & 4,930          & -                 & 4,929           & -                \\
                          & 1                                                                          & 3,075           & 1,025                & 3,075           & 1,025               & 3,075           & 1,025               & 3,075           & 1,025               \\
                          & 2                                                                          & 4,500           & 1,500                & 4,500           & -                & 4,500           & 1,500                & 4,500           & -                \\ \hline
\end{tabular}
\end{adjustbox}
\end{table}

Table~\ref{three sample methods} provides the attack performance of different datasets and GNN models under three sampling methods. From the table, the third sampling method achieves the best average attack accuracy in all dataset and GNN models apart from GIN model on CiteSeer dataset, which obtains the highest average attack accuracy with the first sampling method. From Table~\ref{data points comparison}, we can see that both first and third sampling methods suffer class imbalance, which implies class imbalance can improve the attack accuracy. Besides, the difference in average attack accuracy reaches $7\%$ between the second and third sampling methods on GAT model with DBLP dataset. It indicates that the sampling method influences the attack performance via class imbalance and achieves a maximum gap of $7\%$ for average attack accuracy. Olatunji et al.~\cite{olatunji_membership_2021} used the partially balanced sampling method, while He et al.~\cite{he_node-level_2021} leveraged the random sampling method. We use the balanced sampling method by default to avoid the disturbance brought by the class imbalance. 

\begin{table}[ht]
\centering
\caption{The average attack accuracy of four datasets and GNN models under three sampling methods after ten repetitions.}
\label{three sample methods}
\begin{adjustbox}{max width=0.48\textwidth}
\begin{tabular}{cccccc}
\hline
\begin{tabular}[c]{@{}c@{}}Sampling\\ Method\end{tabular}    & Dataset  & GNN       & \begin{tabular}[c]{@{}c@{}}Test Acc\\ (target model)\end{tabular} & \begin{tabular}[c]{@{}c@{}}Train Acc\\ (target model)\end{tabular} & Avg Acc \\ \hline
\multirow{4}{*}{0} & Cora\_ML & GCN       & 0.68                                                               & 0.851                                                               & 0.621   \\
                   & CiteSeer & GIN       & 0.742                                                              & 0.956                                                               & 0.613   \\
                   & DBLP     & GAT       & 0.653                                                              & 0.682                                                               & 0.614   \\
                   & PubMed   & GraphSAGE & 0.871                                                              & 0.919                                                               & 0.55    \\ \hline
\multirow{4}{*}{1} & Cora\_ML & GCN       & 0.749                                                              & 0.935                                                               & 0.608   \\
                   & CiteSeer & GIN       & 0.775                                                              & 0.927                                                               & 0.591   \\
                   & DBLP     & GAT       & 0.732                                                              & 0.837                                                               & 0.594   \\
                   & PubMed   & GraphSAGE & 0.87                                                               & 0.925                                                               & 0.574   \\ \hline
\multirow{4}{*}{2} & Cora\_ML & GCN       & 0.777                                                              & 0.941                                                               & 0.651   \\
                   & CiteSeer & GIN       & 0.728                                                              & 0.946                                                               & 0.568   \\
                   & DBLP     & GAT       & 0.771                                                              & 0.82                                                                & 0.664   \\
                   & PubMed   & GraphSAGE & 0.866                                                              & 0.919                                                               & 0.581   \\ \hline
\end{tabular}
\end{adjustbox}
\end{table}

\textbf{Attack Model Selection Strategies.} Section~\ref{experiment steps} explains that we select the attack model during the training process based on five metrics, including training accuracy (\textit{train\_acc}), testing accuracy (\textit{test\_acc}), training loss (\textit{train\_loss}), testing loss (\textit{test\_loss}), and evaluation accuracy (\textit{evaluate\_acc}). Table~\ref{selection metrics} gives the average attack accuracy of the selected attack model. From the table, we could observe that selection strategies based on testing accuracy and loss have a slightly better attack accuracy than that based on training accuracy and loss with a maximum gap of $1\%$ average attack accuracy. This paper uses the testing accuracy for selecting the attack model.

Under our environment and settings, previous probability-based methods might not achieve the attack performance reported in their papers~\cite{he_node-level_2021,olatunji_membership_2021}. We attribute this kind of difference to three influence factors analyzed in this section and the change in the attack environment, including the model's architecture, training process, and hyperparameters.

\begin{table*}[ht]
\centering
\caption{The average attack accuracy of four datasets and GNN models with five model selection strategies after ten repetitions.}
\label{selection metrics}
\begin{adjustbox}{max width=0.8\textwidth}
\begin{tabular}{ccccccccc}
\hline
\multirow{2}{*}{Dataset} & \multirow{2}{*}{GNN} & \multirow{2}{*}{\begin{tabular}[c]{@{}c@{}}Test Acc\\ (target model)\end{tabular}} & \multirow{2}{*}{\begin{tabular}[c]{@{}c@{}}Train Acc\\ (target model)\end{tabular}} & \multicolumn{5}{c}{Average accuracy of the attack model selected by different strategies} \\ \cline{5-9} 
                         &                      &                                                                                     &                                                                                      & train\_acc      & test\_acc      & train\_loss      & test\_loss      & evaluate\_acc     \\ \hline
Cora\_ML                 & GCN                  & 0.744                                                                               & 0.943                                                                                & 0.607           & 0.616          & 0.603            & 0.617           & 0.644             \\
CiteSeer                 & GIN                  & 0.771                                                                               & 0.924                                                                                & 0.585           & 0.589          & 0.582            & 0.596           & 0.615             \\
DBLP                     & GAT                  & 0.734                                                                               & 0.837                                                                                & 0.582           & 0.592          & 0.583            & 0.589           & 0.604             \\
PubMed                   & GraphSAGE            & 0.874                                                                               & 0.924                                                                                & 0.572           & 0.578          & 0.565            & 0.589           & 0.604             \\ \hline
\end{tabular}
\end{adjustbox}
\end{table*}

\subsection{Assumptions Relaxation}
\label{assumptions relaxation}
We design and conduct an ablation study for the relaxation of two assumptions about the shadow dataset and the target model's information. In the first experiment, we utilize shadow datasets from other distributions. Secondly, we train the shadow model with different types and architectures from the target model. Finally, we relax those two assumptions together. 

\begin{figure}[t]
\centering
\includegraphics[scale=0.45]{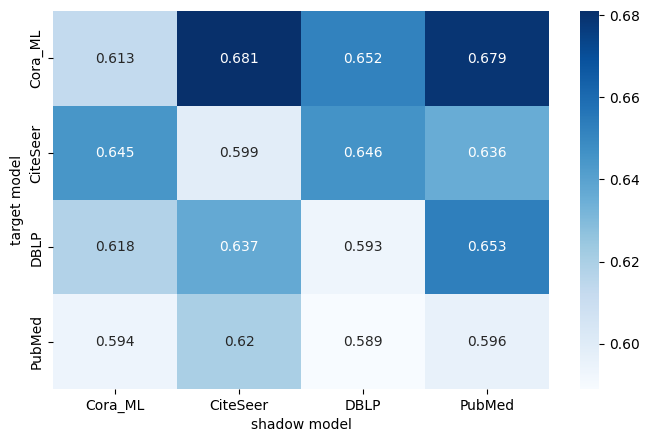}
\caption{The average attack accuracy after ten repetitions while the shadow dataset is from the other distribution. The fixed GNN for shadow and target models is GCN. The x-axis means the setting of the shadow model. The y-axis represents the setting of the target model.}
\label{relaxation_1}
\end{figure}

\begin{figure}[t]
\centering
\includegraphics[scale=0.45]{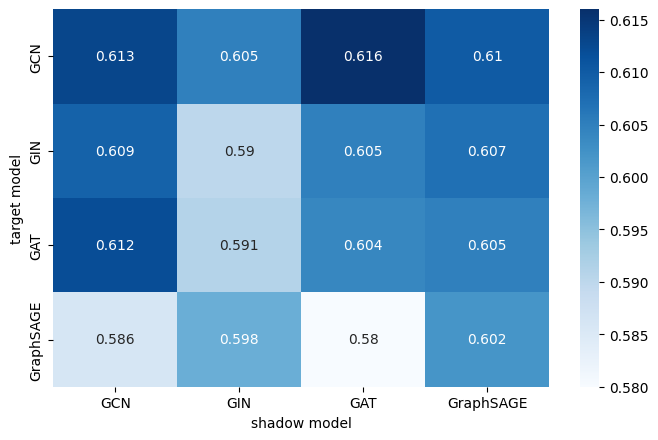}
\caption{The average attack accuracy after ten repetitions while the target model's information is relaxed. The shadow and target datasets are from Cora\_ML.}
\label{relaxation_2}
\end{figure}

\begin{figure}[t]
\centering
\includegraphics[scale=0.45]{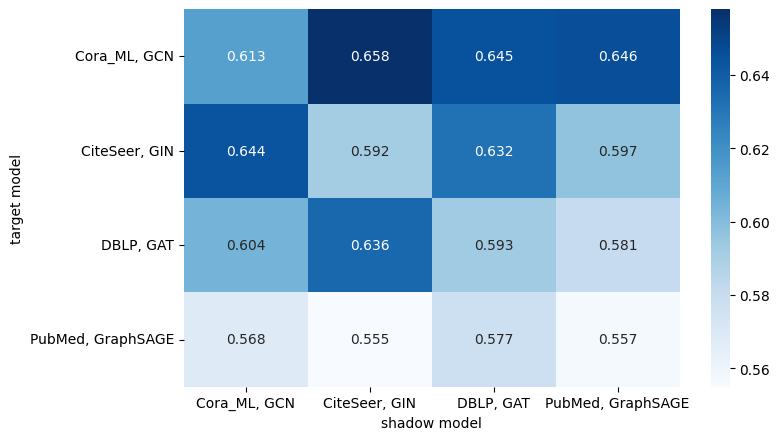}
\caption{The average attack accuracy after ten repetitions while the shadow dataset and the target model's information are relaxed together.}
\label{relaxation_3}
\end{figure}

Figure~\ref{relaxation_1} shows the average attack accuracy comparison of the first experiment. We fix the target model and shadow model to GCN. While the target models are trained with Cora\_ML, CiteSeer, DBLP, and PubMed (each row in the figure), we obtain the largest attack accuracy with shadow datasets from PubMed (0.679), DBLP (0.646), PubMed (0.653), and CiteSeer (0.62), which are not from the same datasets as the target datasets. Figure~\ref{relaxation_2} shows the result of the second experiment. We fix the target dataset and shadow dataset to sample from Cora\_ML. The main reason for selecting Cora\_ML is that the average attack accuracy of Cora\_ML is relatively higher than other datasets, which means the change is evident while relaxing the second assumption. From the figure, we obtain the highest attack accuracy while the target and shadow models are GCN (GAT), GIN (GCN), GAT (GCN), and GraphSAGE (GraphSAGE), most of which do not have the same model type for target and shadow models. Figure~\ref{relaxation_3} illustrates the average attack accuracy of the third experiment. Similarly, the maximum attack accuracy is not from the settings where the target model is trained with the same dataset and model type as the shadow model. From those three figures, we surprisingly find that the relaxation of those two assumptions will increase the attack performance in most cases, which reflects on relatively light colors on the diagonal. Even in the case where the attack performance decreases, the reduction degree is low. Therefore, the attack performance primarily increases while relaxing the assumptions about the shadow dataset and target model's information. The possible reason for this phenomenon is that the attack features of members (or non-members) are similar or are converted to be alike in the attack model under settings where the shadow model's distribution and the target model's information are different. This phenomenon is also exposed in previous works~\cite{Salem_Membership_2019,he_node-level_2021}.

\subsection{Defenses}
\label{defenses_1}

Table~\ref{defenses} shows the average attack accuracy of ten repetitions against four different defenses. The defenses include Normalization (the BatchNorm), Dropout (0.5), Regularization (the Adam with weight decay of 0.5), and Jumping knowledge (concatenation). Here, we select GCN model on Cora\_ML dataset with a high overfitting level to expose the influence of defenses due to its relatively high attack performance. From the table, we can observe that the average accuracy decreases when applying Normalization (row 1, 0.586) or Regularization (row 3, 0.573) compared with no defenses (row 0, 0.596). However, the average accuracy slightly increases when only using the Dropout (row 2, 0.602) or Jumping knowledge (row 4, 0.606). Besides, the combinations between four defenses could raise (row 5, 7, 9, 11, 12, 13, 15) or lower (row 6, 8, 10, 14) the average accuracy. In addition, Regularization is in four rows (6, 8, 10, 14), where the average accuracy is decreased. It does not mean the average accuracy will be lower with Regularization combined with other defenses, like rows 11 and 13. The results show that Regularization and Normalization could slightly decrease the average attack accuracy while applying alone. Moreover, those four defenses cannot prevent our label-only MIA completely with higher average accuracy than 0.5 apart from row 14.

\begin{table}[ht]
\centering
\caption{The average attack accuracy after ten repetitions under different defenses (Cora\_ML, GCN).}
\label{defenses}
\begin{adjustbox}{max width=0.48\textwidth}
\begin{tabular}{cccccc}
\hline
Row & Normalization & Dropout & Regularization & \begin{tabular}[c]{@{}c@{}}Jumping\\ Knowledge\end{tabular} & \begin{tabular}[c]{@{}c@{}}Avg Acc\end{tabular} \\ \hline
0 &×             & ×       & ×              & ×                                                           & 0.596                                                                                     \\
1 &$\surd$             & ×       & ×              & ×                                                           & 0.586                                                                                     \\
2 &×             & $\surd$       & ×              & ×                                                           & 0.602                                                                                     \\
3 &×             & ×       & $\surd$              & ×                                                           & 0.573                                                                                     \\
4 &×             & ×       & ×              & $\surd$                                                           & 0.606                                                                                     \\
5 &$\surd$             & $\surd$       & ×              & ×                                                           & 0.609                                                                                     \\
6 &$\surd$             & ×       & $\surd$              & ×                                                           & 0.581                                                                                     \\
7 &$\surd$             & ×       & ×              & $\surd$                                                           & 0.615                                                                                     \\
8 &×             & $\surd$       & $\surd$              & ×                                                           & 0.55                                                                                      \\
9 &×             & $\surd$       & ×              & $\surd$                                                           & 0.607                                                                                     \\
10 &×             & ×       & $\surd$              & $\surd$                                                           & 0.52                                                                                      \\
11 &$\surd$             & $\surd$      & $\surd$              & ×                                                           & 0.609                                                                                     \\
12 &$\surd$             & $\surd$       & ×              & $\surd$                                                           & 0.614                                                                                     \\
13 &$\surd$             & ×       & $\surd$              & $\surd$                                                           & 0.618                                                                                     \\
14 &×             & $\surd$       & $\surd$              & $\surd$                                                           & 0.497                                                                                     \\
15 &$\surd$             & $\surd$       & $\surd$              & $\surd$                                                           & 0.613                                                                                     \\ \hline
\end{tabular}
\end{adjustbox}
\end{table}

\section{Related Work}
\label{sec:related}
\textbf{Membership Inference Attacks.} In 2016, Shokri et al.~\cite{shokri_membership_2017} proposed MIAs to machine learning models by training the attack model with attack features through the shadow model and dataset. Then, Salem et al.~\cite{Salem_Membership_2019} relaxed some assumptions from~\cite{shokri_membership_2017} and achieved competitive attack performance. Recently, researchers have been paying more attention to MIAs. At the same time, membership inference attacks widely spread to multiple data types (numerical data, image, text, and graph) and model categories (classical algorithm, CNN, RNN, GAN, and GNN)~\cite{wu_adapting_2021,Ruiz_classical_2022,hayes_logan_2019,zhang22usenix,song_auditing_2019}. The critical point of a successful attack is distinguishing between the training and testing data. Previous research suggested feasible keystone metrics, like prediction probability vector and loss~\cite{yeom_privacy_2018}. The intuition is that the model is more confident in training data points than testing data points because the model is overfitting. Overfitting is the most acceptable reason for MIA but not the only factor~\cite{yeom_privacy_2018,Truex_Demystify_2018}.

\textbf{Membership Inference Attacks against GNNs.} Wu et al.~\cite{wu_adapting_2021} firstly proposed MIAs to GNNs for classification tasks. Their attack methods utilized the prediction probability vector and some metrics computed based on it. In contrast, He et al.~\cite{he_node-level_2021} concentrated on MIAs against node-level GNNs. They detected the node's existence with its top two probability scores by feeding it into the model with its 0-hop or 2-hop neighborhood nodes. Olatunji et al.~\cite{olatunji_membership_2021} put the overall training or testing subgraph into the model to simultaneously obtain nodes' prediction probability vectors. Those three methods are related to the prediction probability vector, which is infeasible under the label-only condition with just a prediction label as the output. This paper presents a solution for attacking GNNs without the prediction probability vector.

\textbf{Label-only Membership Inference Attacks.} Li et al.~\cite{li_label-leaks_2020} put forward two strategies to attack traditional neural networks under label-only condition. The first method trains a relabelled shadow model for obtaining the prediction probability vector of data points in the target dataset. The second method measures the distance between the data point and its adversarial example. Then, distinguish members and non-members by comparing this distance with the threshold. In the meantime, Choquette-Choo et al.~\cite{choquette-choo_label-only_2021} put forward two approaches for MIAs under a label-only environment. One method leverages the prediction correctness situation of a data point and its augmented versions for membership prediction. The other method is similar to the second approach proposed by Li et al.. Those two works inspire our research of label-only MIAs against GNNs. In addition, Rahimian et al.~\cite{Rahimian_Sampling_2021} constructed the posterior vector of the data point by feeding multiple perturbed versions to the target model and analyzing the predictions of those perturbed data of this data point. 


\section{Conclusions and Future Work}
In this paper, we propose a method of implementing MIA against GNNs under the label-only condition. The average attack accuracy, precision, and AUC values of our label-only MIA are around 0.6 in most experiments, which are competitive or even better than previous probability-based MIAs implemented in the same environment and settings. In addition, we explore the influence factors of our label-only MIA. The higher overfitting level impacts the attack performance, i.e., it could increase or decrease our label-only MIA's attack accuracy. The sampling method also influences the attack performance and achieves a maximum gap of $7\%$ on average attack accuracy. Model selection strategies with testing accuracy and loss have a slightly better attack accuracy than training accuracy and loss with a maximum gap of $1\%$ on average attack accuracy. Then, we consider scenarios where two assumptions about the shadow dataset and the target model's information are relaxed. Surprisingly, the relaxation of those two assumptions will increase the attack performance in most experiments. Finally, we explore label-only MIA against four defense methods and their combinations. The results show that those four defenses cannot prevent our label-only MIA completely. This paper only focuses on node classification tasks. We leave label-only MIA against graph-level GNNs as feature work.



\bibliographystyle{ACM-Reference-Format}
\bibliography{paper}


\appendix

\section{Additional Results}
\label{table_explanation}

Table~\ref{previous MIA performance} indicates the attack performance of four previous probability-based MIAs, including 0-hop, 2-hop, the combination of 0-hop and 2-hop, and all probability methods. We implement those four MIAs under the same settings as the results of Table~\ref{label-only MIA performance}. The settings include the same target and shadow models (a low overfitting level) trained with the same dataset split, the training of the attack models, and the model selection strategy. Each result row in Table~\ref{previous MIA performance} corresponds with the result row under the same dataset and GNN model in Table~\ref{label-only MIA performance}. For each row, the dataset and GNN are used to train the target and shadow models. The "0-hop", "2-hop", "o-hop and 2-hop combination", and "all probability" in the table indicate the results of four probability-based MIAs. Each result has six values: the average accuracy, precision, recall, AUC value, F1 score, and TPR under low FPR (0.01) after ten repetitions. We highlight the average accuracy of each result for a clear comparison with the red color.

\begin{table*}[ht]
\centering
\caption{The attack performance of probability-based methods after ten repetitions.}
\label{previous MIA performance}
\begin{adjustbox}{max width=1.048\textwidth}
\begin{tabular}{|c|c|cccc|lll}
\cline{1-6}
\multirow{2}{*}{Dataset}  & \multirow{2}{*}{GNN} & \multicolumn{4}{c|}{probability-based methods (avg acc, pre, rec, auc, f1, low\_fpr\_0.01\_tpr)}                                                                                                                                                                 & \multicolumn{1}{c}{} & \multicolumn{1}{c}{} & \multicolumn{1}{c}{} \\ \cline{3-6}
                          &                      & \multicolumn{1}{c|}{0-hop}                                          & \multicolumn{1}{c|}{2-hop}                                          & \multicolumn{1}{c|}{0-hop and 2-hop combination}                    & all probability                                &                      &                      &                      \\ \cline{1-6}
\multirow{4}{*}{Cora\_ML} & GAT                  & \multicolumn{1}{c|}{{[}\textcolor{red}{0.576}, 0.552, 0.634, 0.626, 0.546, 0.076{]}} & \multicolumn{1}{c|}{{[}\textcolor{red}{0.549}, 0.508, 0.457, 0.61, 0.409, 0.033{]}}  & \multicolumn{1}{c|}{{[}\textcolor{red}{0.588}, 0.563, 0.62, 0.629, 0.548, 0.088{]}}  & {[}\textcolor{red}{0.571}, 0.511, 0.493, 0.652, 0.454, 0.078{]} &                      &                      &                      \\ \cline{2-6}
                          & GCN                  & \multicolumn{1}{c|}{{[}\textcolor{red}{0.691}, 0.67, 0.874, 0.765, 0.746, 0.157{]}}  & \multicolumn{1}{c|}{{[}\textcolor{red}{0.633}, 0.622, 0.807, 0.685, 0.687, 0.059{]}} & \multicolumn{1}{c|}{{[}\textcolor{red}{0.693}, 0.675, 0.851, 0.765, 0.739, 0.142{]}} & {[}\textcolor{red}{0.635}, 0.681, 0.639, 0.723, 0.591, 0.121{]} &                      &                      &                      \\ \cline{2-6}
                          & GIN                  & \multicolumn{1}{c|}{{[}\textcolor{red}{0.613}, 0.611, 0.761, 0.692, 0.655, 0.092{]}} & \multicolumn{1}{c|}{{[}\textcolor{red}{0.575}, 0.577, 0.722, 0.604, 0.609, 0.009{]}} & \multicolumn{1}{c|}{{[}\textcolor{red}{0.6}, 0.58, 0.724, 0.666, 0.629, 0.106{]}}    & {[}\textcolor{red}{0.55}, 0.587, 0.52, 0.575, 0.51, 0.026{]}    &                      &                      &                      \\ \cline{2-6}
                          & GraphSAGE            & \multicolumn{1}{c|}{{[}\textcolor{red}{0.627}, 0.68, 0.732, 0.747, 0.636, 0.136{]}}  & \multicolumn{1}{c|}{{[}\textcolor{red}{0.626}, 0.685, 0.73, 0.762, 0.625, 0.107{]}}  & \multicolumn{1}{c|}{{[}\textcolor{red}{0.639}, 0.66, 0.793, 0.74, 0.68, 0.124{]}}    & {[}\textcolor{red}{0.61}, 0.66, 0.324, 0.637, 0.372, 0.115{]}   &                      &                      &                      \\ \cline{1-6}
\multirow{4}{*}{CiteSeer} & GAT                  & \multicolumn{1}{c|}{{[}\textcolor{red}{0.533}, 0.478, 0.329, 0.562, 0.338, 0.013{]}} & \multicolumn{1}{c|}{{[}\textcolor{red}{0.518}, 0.422, 0.436, 0.539, 0.399, 0.009{]}} & \multicolumn{1}{c|}{{[}\textcolor{red}{0.524}, 0.426, 0.467, 0.555, 0.414, 0.014{]}} & {[}\textcolor{red}{0.515}, 0.38, 0.281, 0.531, 0.295, 0.01{]}   &                      &                      &                      \\ \cline{2-6}
                          & GCN                  & \multicolumn{1}{c|}{{[}\textcolor{red}{0.555}, 0.456, 0.45, 0.567, 0.439, 0.011{]}}  & \multicolumn{1}{c|}{{[}\textcolor{red}{0.559}, 0.523, 0.475, 0.6, 0.463, 0.017{]}}   & \multicolumn{1}{c|}{{[}\textcolor{red}{0.551}, 0.526, 0.425, 0.575, 0.443, 0.009{]}} & {[}\textcolor{red}{0.528}, 0.507, 0.366, 0.525, 0.388, 0.007{]} &                      &                      &                      \\ \cline{2-6}
                          & GIN                  & \multicolumn{1}{c|}{{[}\textcolor{red}{0.535}, 0.48, 0.618, 0.562, 0.508, 0.01{]}}   & \multicolumn{1}{c|}{{[}\textcolor{red}{0.541}, 0.535, 0.759, 0.567, 0.601, 0.011{]}} & \multicolumn{1}{c|}{{[}\textcolor{red}{0.534}, 0.534, 0.728, 0.568, 0.581, 0.01{]}}  & {[}\textcolor{red}{0.515}, 0.498, 0.248, 0.519, 0.266, 0.01{]}  &                      &                      &                      \\ \cline{2-6}
                          & GraphSAGE            & \multicolumn{1}{c|}{{[}\textcolor{red}{0.522}, 0.476, 0.411, 0.534, 0.348, 0.014{]}} & \multicolumn{1}{c|}{{[}\textcolor{red}{0.54}, 0.416, 0.381, 0.551, 0.354, 0.012{]}}  & \multicolumn{1}{c|}{{[}\textcolor{red}{0.53}, 0.417, 0.393, 0.569, 0.364, 0.011{]}}  & {[}\textcolor{red}{0.519}, 0.312, 0.26, 0.523, 0.268, 0.01{]}   &                      &                      &                      \\ \cline{1-6}
\multirow{4}{*}{DBLP}     & GAT                  & \multicolumn{1}{c|}{{[}\textcolor{red}{0.516}, 0.539, 0.175, 0.547, 0.231, 0.011{]}} & \multicolumn{1}{c|}{{[}\textcolor{red}{0.514}, 0.471, 0.276, 0.539, 0.278, 0.011{]}} & \multicolumn{1}{c|}{{[}\textcolor{red}{0.517}, 0.479, 0.274, 0.54, 0.299, 0.01{]}}   & {[}\textcolor{red}{0.503}, 0.52, 0.161, 0.521, 0.188, 0.015{]}  &                      &                      &                      \\ \cline{2-6}
                          & GCN                  & \multicolumn{1}{c|}{{[}\textcolor{red}{0.532}, 0.545, 0.406, 0.561, 0.415, 0.011{]}} & \multicolumn{1}{c|}{{[}\textcolor{red}{0.534}, 0.548, 0.424, 0.556, 0.419, 0.013{]}} & \multicolumn{1}{c|}{{[}\textcolor{red}{0.532}, 0.543, 0.412, 0.562, 0.421, 0.011{]}} & {[}\textcolor{red}{0.527}, 0.524, 0.416, 0.549, 0.432, 0.014{]} &                      &                      &                      \\ \cline{2-6}
                          & GIN                  & \multicolumn{1}{c|}{{[}\textcolor{red}{0.514}, 0.436, 0.325, 0.531, 0.299, 0.012{]}} & \multicolumn{1}{c|}{{[}\textcolor{red}{0.517}, 0.499, 0.373, 0.529, 0.339, 0.006{]}} & \multicolumn{1}{c|}{{[}\textcolor{red}{0.524}, 0.453, 0.358, 0.538, 0.318, 0.006{]}} & {[}\textcolor{red}{0.507}, 0.372, 0.277, 0.527, 0.266, 0.006{]} &                      &                      &                      \\ \cline{2-6}
                          & GraphSAGE            & \multicolumn{1}{c|}{{[}\textcolor{red}{0.531}, 0.53, 0.49, 0.553, 0.465, 0.011{]}}   & \multicolumn{1}{c|}{{[}\textcolor{red}{0.538}, 0.537, 0.474, 0.558, 0.463, 0.012{]}} & \multicolumn{1}{c|}{{[}\textcolor{red}{0.539}, 0.548, 0.391, 0.562, 0.423, 0.015{]}} & {[}\textcolor{red}{0.554}, 0.557, 0.576, 0.583, 0.536, 0.017{]} &                      &                      &                      \\ \cline{1-6}
\multirow{4}{*}{PubMed}   & GAT                  & \multicolumn{1}{c|}{{[}\textcolor{red}{0.498}, 0.415, 0.185, 0.503, 0.198, 0.008{]}} & \multicolumn{1}{c|}{{[}\textcolor{red}{0.5}, 0.341, 0.384, 0.497, 0.306, 0.009{]}}   & \multicolumn{1}{c|}{{[}\textcolor{red}{0.497}, 0.442, 0.365, 0.501, 0.352, 0.009{]}} & {[}\textcolor{red}{0.505}, 0.464, 0.531, 0.511, 0.456, 0.01{]}  &                      &                      &                      \\ \cline{2-6}
                          & GCN                  & \multicolumn{1}{c|}{{[}\textcolor{red}{0.506}, 0.353, 0.217, 0.515, 0.229, 0.009{]}} & \multicolumn{1}{c|}{{[}\textcolor{red}{0.506}, 0.478, 0.291, 0.512, 0.286, 0.009{]}} & \multicolumn{1}{c|}{{[}\textcolor{red}{0.504}, 0.446, 0.203, 0.508, 0.224, 0.008{]}} & {[}\textcolor{red}{0.507}, 0.476, 0.342, 0.516, 0.353, 0.007{]} &                      &                      &                      \\ \cline{2-6}
                          & GIN                  & \multicolumn{1}{c|}{{[}\textcolor{red}{0.502}, 0.393, 0.649, 0.508, 0.455, 0.007{]}} & \multicolumn{1}{c|}{{[}\textcolor{red}{0.502}, 0.398, 0.376, 0.508, 0.294, 0.007{]}} & \multicolumn{1}{c|}{{[}\textcolor{red}{0.501}, 0.267, 0.375, 0.504, 0.262, 0.009{]}} & {[}\textcolor{red}{0.499}, 0.239, 0.125, 0.497, 0.134, 0.006{]} &                      &                      &                      \\ \cline{2-6}
                          & GraphSAGE            & \multicolumn{1}{c|}{{[}\textcolor{red}{0.504}, 0.404, 0.273, 0.513, 0.307, 0.008{]}} & \multicolumn{1}{c|}{{[}\textcolor{red}{0.508}, 0.403, 0.295, 0.521, 0.302, 0.009{]}} & \multicolumn{1}{c|}{{[}\textcolor{red}{0.504}, 0.451, 0.293, 0.52, 0.338, 0.009{]}}  & {[}\textcolor{red}{0.511}, 0.416, 0.334, 0.525, 0.351, 0.008{]} &                      &                      &                      \\ \cline{1-6}
\end{tabular}
\end{adjustbox}
\end{table*}



\end{document}